\documentclass[aps,reprint]{revtex4-2}
\usepackage{bm, amsmath, amsfonts, mathtools, physics}

\usepackage{color}
\usepackage{hyperref}
\usepackage{float}
\hypersetup{colorlinks=true,linkcolor=blue, citecolor=blue}

\usepackage{orcidlink}

\begin{document}

    \title{Self-Organized Dynamic Reactors: \\
   Product–Condensate Feedback Accelerates Reactions}
    
    \author{Ryosuke Nishide\orcidlink{0000-0003-2586-5177}}
    \email{ryosuke.nishide@nbi.ku.dk}

    \author{Kunihiko Kaneko\orcidlink{0000-0001-6400-8587}}

    \affiliation{
    Niels Bohr Institute, University of Copenhagen, Jagtvej 155 A, Copenhagen N, 2200, Denmark}

\begin{abstract}
Condensates can accelerate reactions by concentrating reactants, yet product accumulation can eliminate this advantage. 
We show that product-induced suppression of condensation generates rotating, deforming, and propagating condensate domains through spontaneous symmetry breaking. 
Their reorganization shifts reaction sites away from accumulated products while retaining reactant enrichment, yielding rates above those of static condensates and homogeneous states. 
Reaction flows can thus self-organize condensates into dynamic reactors, with implications for biological and synthetic systems.
\end{abstract}

\maketitle
Life is sustained by nonequilibrium reaction flows in which substrates are supplied, transformed, and their products undergo subsequent reactions.
Biomolecular condensates spatially organize these flows, enriching enzymes and substrates where reactions proceed and regulating biochemical function \cite{hyman2014liquidliquid, banani2017biomolecular, lyon2021aframework, aierken2026roadmap}.
At the origin of life, molecular compartments such as coacervates could have concentrated molecules and sustained prebiotic reaction networks \cite{oparin1938theorigin, koga2011peptidenucleotide, drobot2018compartmentalised, nakashima2021active}.

Condensates can accelerate reactions relative to homogeneous conditions by spatially concentrating their reactants \cite{kuffner2020acceleration, peeples2021mechanistic, gilgarcia2024local}.
This enrichment, however, does not necessarily lead to faster reactions: the condensed environment also alters catalytic kinetics and molecular mobility and can even suppress production \cite{peeples2021mechanistic, harris2024regulation, schmit2024physical}.
Moreover, product transport and subsequent reactions influence the overall rate, requiring consideration of the entire reaction flow rather than production alone.
In particular, when the product is removed from the condensate more slowly than it is formed \cite{lye1999application, smokers2024how}, it accumulates locally and inhibits further production, yet this effect has received little attention.
We therefore investigate how product inhibition affects the overall reaction rate and whether condensate self-organization can enhance it by reducing this inhibition.

To address this question, we focus on product-dependent suppression of condensation as a mechanism for self-regulation.
Both small molecules such as ATP and scaffold-binding ligands \cite{patel2017atp, hernandezcandia2021amodular} and reactions that modify condensate components \cite{donau2020active, wurtz2018chemical} can destabilize condensates, providing plausible mechanisms for such regulation.
This suppression, however, entails a trade-off: while it may reduce product accumulation, it also reduces reactant enrichment, the source of the condensate’s advantage. 
It is therefore nontrivial whether this mechanism enhances the reaction.

Here, we introduce a minimal model in which the reaction product itself suppresses the intermolecular interactions that stabilize the condensate. 
We show that rotating and propagating condensate domains emerge through spontaneous symmetry breaking.
At intermediate suppression strengths, these dynamic domains reduce product accumulation while retaining reactant enrichment, yielding higher reaction rates than both a static condensate and the homogeneous state.

We describe the condensate-forming component by a conserved volume fraction $\phi(\bm{x},t)$ with spatial mean $\phi_0$ and denote the product concentration by $p(\bm{x},t)$.
Reactants are enriched in $\phi$-rich regions, where the product is preferentially generated, whereas its subsequent conversion occurs mainly in the surrounding $\phi$-poor region [Fig.~\ref{fig5}(a)].
We denote the local rates of product formation and conversion by $r_{\rm prod}(\phi,p)$ and $r_{\rm conv}(\phi,p)$, respectively.
Condensation enhances production, whereas product accumulation suppresses it, so that 
$\partial_\phi r_{\rm prod}>0$ and $\partial_p r_{\rm prod}<0$ [Fig.~\ref{fig5}(b)]. 
The conversion rate increases with product concentration but decreases with $\phi$, giving 
$\partial_p r_{\rm conv}>0$ and $\partial_\phi r_{\rm conv}<0$ [Fig.~\ref{fig5}(b)].

\begin{figure*}[t]
    \includegraphics[keepaspectratio,scale=1.0]{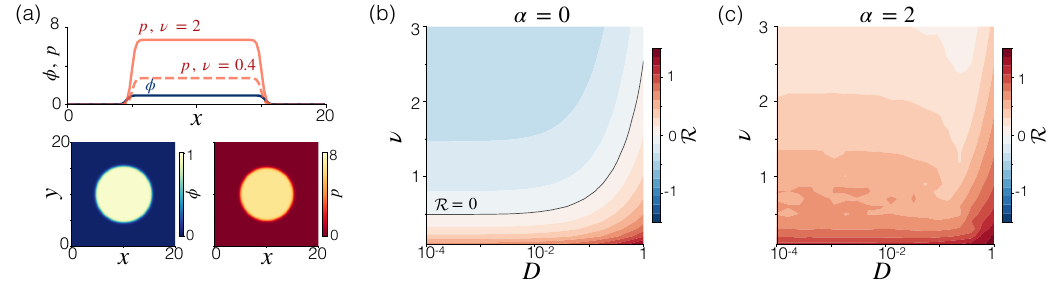}
    \caption{Condensate and product profiles and relative reaction rate $\mathcal R$.
    (a) Static profiles of the condensate fraction $\phi$ (blue) and product concentration $p$ (red) at $\alpha = 0$ and $D=10^{-3}$
    Top: cross sections at $y= 10$ for $\nu=0.4$ (dashed) and $2$ (solid). 
    Bottom: $\phi$ (left) and $p$ (right) in the $(x,y)$ plane at $\nu=2$.
    (b), (c) $\mathcal R$ in the $(D,\nu)$ plane at $\alpha=0$ in (b) and $\alpha=2$ in (c).}
\label{fig1}
\end{figure*}

As a minimal biochemical scheme consistent with these relations, we consider $S+E\rightarrow P+E$ within the condensate and $P\rightarrow\emptyset$ in the dilute phase [Fig.~\ref{fig5}(c)].
We assume that the substrate is continuously supplied and rapidly partitions into the condensate, while the enzyme either partitions similarly or itself constitutes the condensate-forming component.
Their local concentrations then satisfy $[S]\simeq S_0\phi$ and $[E]\simeq E_0\phi$, where $S_0$ and $E_0$ are constant.
Thus, $r_{\rm prod}$ is proportional to $\phi^2$, while local product accumulation suppresses it through product inhibition (see Supplemental Material \cite{supp}). 
Accordingly, the local reaction rates are
\begin{equation}
r_{\rm prod} = \eta\frac{\phi^2}{1+p/K_P},
\quad
r_{\rm conv} = \lambda(1-\phi)p,
\label{eq:reaction_rates}
\end{equation}
where $\eta$ and $\lambda$ are the production and conversion rate constants, respectively, and $K_P$ is the product-inhibition constant.
We use their ratio, $\nu\equiv\eta/\lambda$, as an effective production strength.

We describe the thermodynamics by the dimensionless free-energy functional
\begin{align}\label{eq:free_energy}
    F=\int d\bm{x}\left[ f_{\phi}(\phi)+f_{p}(p)+\frac{\chi(p)}{2}\phi^2+\frac{\kappa}{2}|\nabla\phi|^2\right],
\end{align}
where $f_{\phi}=\phi \ln\phi+(1-\phi)\ln(1-\phi)$, $f_{p}=p( \ln p-1)$, and $\kappa>0$.
Here, $f_{\phi}$ is the lattice-mixing contribution of the condensate-forming component and an implicit solvent, while $f_p$ treats the product as a dilute ideal solute.
We introduce product--condensate coupling through $\chi(p)$, assuming that product suppresses the attraction driving condensation, $\chi'(p)>0$. 
We use the simplest form $\chi(p)=\chi_0+\alpha p$, where $\chi_0<0$ is the interaction strength in the absence of product and $\alpha>0$ is the coupling strength.
The same form arises to leading order upon eliminating a rapidly relaxing intermediate chemical species \cite{supp}.
Thus, product accumulation makes $\chi(p)$ less negative and locally suppresses condensation.
At $p=0$, we assume that $\phi$ undergoes spinodal decomposition, $f_\phi''(\phi_0)+\chi_0<0$.

Combining chemical-potential-driven transport with reactions, we obtain
\begin{align}
    \partial_t\phi
    &=\nabla^2\left[\ln\{\phi/(1-\phi)\}
    +(\chi_0+\alpha p)\phi
    -\kappa\nabla^2\phi\right],\label{eq:phi_dynamics}\\
    \partial_tp
    &=D\left[\nabla^2 p+\alpha \nabla\cdot(p\phi\nabla\phi)\right]
    +r_{\mathrm{prod}}-r_{\mathrm{conv}},\label{eq:p_dynamics}
\end{align}
where $D$ is the dimensionless product diffusivity.
Here, time and the reaction rates are scaled by the mobility of $\phi$.
The derivation is given in the Supplemental Material~\cite{supp}.
While transport follows thermodynamics, the reactions drive the system out of equilibrium.
The product-dependent term $\alpha p\phi$ in Eq.~(\ref{eq:phi_dynamics}) locally suppresses condensation where product accumulates.
The same coupling also generates the product flux $\bm J_p^\alpha=-D\alpha p\phi\nabla\phi$ toward lower $\phi$.
At $D=0$, both product diffusion and the flux vanish, and Eq.~(\ref{eq:p_dynamics}) reduces to 
\begin{align}
    \partial_t p=r_{\rm prod}-r_{\rm conv}.\label{eq:p_dynamics_without_D}
\end{align}
Because the term $\alpha p\phi$ remains in
Eq.~(\ref{eq:phi_dynamics}), this limit allows the effect of the product on $\phi$ dynamics to be examined independently of $\bm J_p^\alpha$.

\begin{figure*}[t]
    \includegraphics[keepaspectratio,scale=1.0]{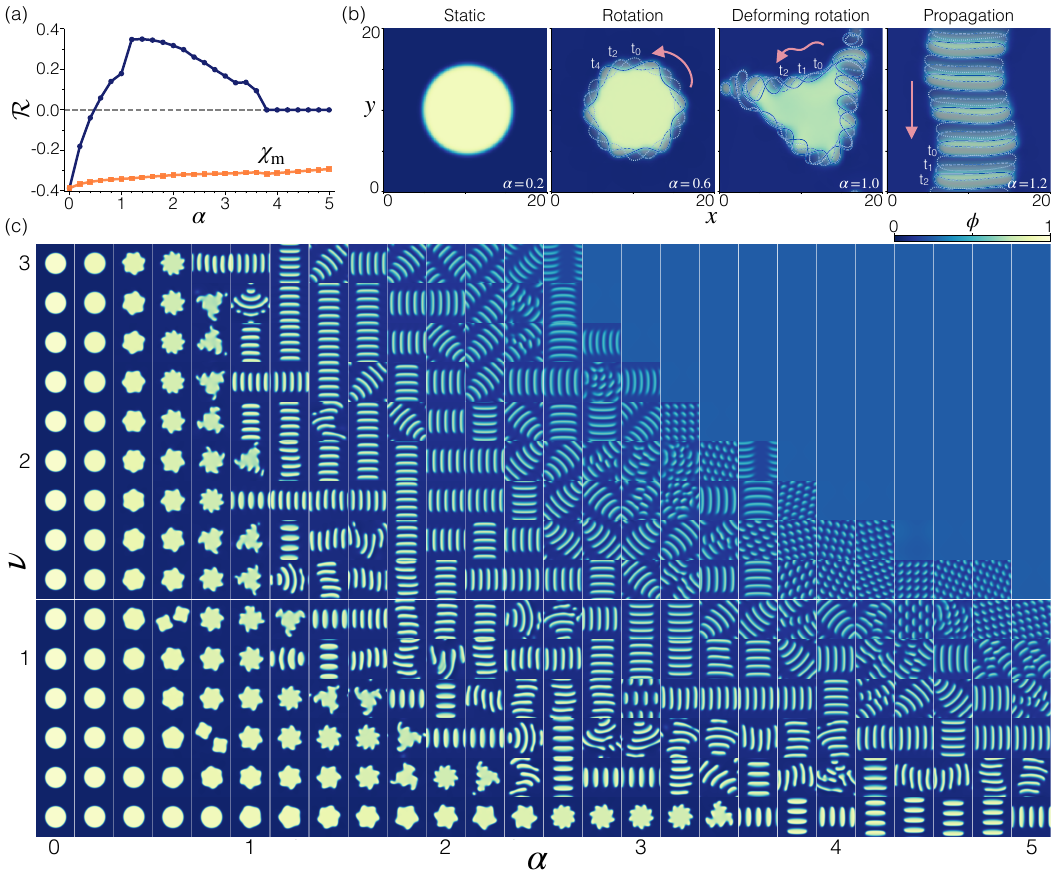}
    \caption{Product-induced suppression of condensation generates dynamic domains and enhances reaction.
    (a) $\mathcal R$ against $\alpha$ for $\chi(p)$ (blue) and $\chi_m$ (orange) at $D=10^{-3}$ and $\nu=2$.
    Simulations are evolved to $t_0=2000$, and $\mathcal R$ is averaged over $t_0\leq t\leq t_0+100$.
    (b) Condensate profiles for the static ($\alpha=0.2$), rotating ($\alpha=0.6$), deforming-rotation ($\alpha=1.0$), and propagating ($\alpha=1.2$) states at $D=10^{-3}$ and $\nu=2$.
    For the dynamic states, profiles at three labeled times are overlaid, with the $\phi=0.5$ contours shown as dotted, dashed, and solid lines, where $t_n=t_0+10n$ ($n\in\mathbb Z$).
    (c) Condensate profiles in the $(\alpha,\nu)$ plane at $D=10^{-3}$.
    The corresponding product profiles are shown in Fig.~S1(b); see also Movie~1 for the dynamics~\cite{supp}.}
    \label{fig2}
\end{figure*}

To quantify reaction enhancement, we calculate the spatially averaged production and subsequent conversion rates,
$R_{\rm prod}=\langle r_{\rm prod}\rangle$ and
$R_{\rm conv}=\langle r_{\rm conv}\rangle$, respectively.
Here, $\langle\cdot\rangle$ denotes a spatial average.
Spatially averaging Eq.~(\ref{eq:p_dynamics}) gives 
$\partial_t\langle p\rangle=R_{\rm prod}-R_{\rm conv}$.
Taking a long-time average, denoted by an overbar, then gives
$\overline R_{\rm prod}=\overline R_{\rm conv}$.
We measure reaction enhancement relative to the homogeneous state by
\begin{equation}
\mathcal R
=\frac{\overline R_{\rm prod}}{R_{\rm hom}}-1
=\frac{\overline R_{\rm conv}}{R_{\rm hom}}-1,
\label{eq}
\end{equation}
where $R_{\rm hom}$ is the reaction rate in the homogeneous state $(\phi,p)=(\phi_0,p_0)$ (see Appendix).
Thus, $\mathcal R>0$ ($<0$) indicates a higher (lower) reaction rate than in the homogeneous state.

Within this model, $\mathcal R$ can become negative only through product inhibition.
In the no-inhibition limit $K_P\rightarrow\infty$, production becomes independent of $p$, with $r_{\rm prod}=\eta\phi^2$.
The long-time averaged rate then satisfies
\begin{equation}
\overline R_{\rm prod}=\overline R_{\rm conv}
=\eta\overline{\langle\phi^2\rangle}
\geq
\eta\phi_0^2=R_{\rm hom}.
\end{equation}
Thus, $\mathcal R\geq0$ holds in the absence of product inhibition.

We perform simulations in two dimensions with periodic boundary conditions.
Unless stated otherwise, we set $\phi_0=0.25$, $\chi_0=-7$, $\kappa=0.2$, and $\lambda=2$.
Numerical details are given in the Supplemental Material \cite{supp}.

We first set $\alpha=0$ and calculate $\mathcal R$ over the $(D,\nu)$ plane \footnote{$\lambda$ and $K_P$
can be absorbed into rescaled $D$ and $\nu$ \cite{supp}.}.
The dynamics of $\phi$ decouples from $p$, so the stationary $\phi$ profile is independent of $D$ and $\nu$, while $p$ localizes within the condensate [Fig.~\ref{fig1}(a)].
Figure~\ref{fig1}(b) shows that $\mathcal R$ decreases as $\nu$ increases and becomes negative at sufficiently large $\nu$.
As $D$ decreases, the $\mathcal R=0$ boundary shifts toward smaller $\nu$ and approaches $\nu\simeq0.5$.
At $D=10^{-3}$, the profiles for $\nu=0.4$ and $2$ show that the product concentration within the condensate is higher at larger $\nu$ [Fig.~\ref{fig1}(a)]. 
Thus, slow diffusion causes product accumulation within the condensate, and the resulting inhibition can reduce the reaction rate below that in the homogeneous state.

We next include product-dependent suppression of condensation by setting $\alpha=2$ and calculate $\mathcal R$ over the same $(D,\nu)$ plane. 
Although $\mathcal R$ still decreases as $\nu$ increases, it remains nonnegative throughout the plane [Fig.~\ref{fig1}(c)]. 
Moreover, $\mathcal R$ exceeds its value at $\alpha=0$ throughout the plane, including the region where condensation already enhances the reaction. 
Product-dependent suppression thus eliminates the $\mathcal R<0$ region and further increases $\mathcal R$ where it is already positive.

To understand why product-dependent suppression of condensation increases $\mathcal R$, we investigate how $\mathcal R$ and the spatial patterns change with the suppression strength $\alpha$ (Figs.~\ref{fig2} and S1~\cite{supp}).
To focus on a regime where product inhibition is pronounced at $\alpha=0$, we set $D=10^{-3}$ and first fix $\nu=2$.
At $\alpha=0$, the condensate is static and $\mathcal R<0$. 
Increasing $\alpha$ to $0.2$ raises $\mathcal R$, although it remains negative, and the condensate remains static. 
As $\alpha$ is increased to $0.4$, $\mathcal R$ rises further but remains negative, while the circular condensate becomes star-shaped and rotating.
With a further increase to $\alpha=0.6$, the deformation grows and the rotation persists, while $\mathcal R$ becomes positive. 
At $\alpha=1.2$, a propagating pattern with multiple domains appears, and $\mathcal R$ reaches its maximum. 
For larger $\alpha$, propagation persists with oscillations and deformation. 
Although $\mathcal R$ remains positive, it gradually decreases together with the maximum value of $\phi$. 
Above $\alpha=3.8$, the patterns disappear, the system becomes homogeneous, and $\mathcal R$ returns to zero.

Remarkably, product-dependent suppression therefore does not simply change the condensate profile but generates sustained rotation and propagation. 
Since the system is spatially symmetric, these dynamics arise through spontaneous symmetry breaking.
The sign change and maximum of $\mathcal R$ both occur within these dynamic regimes, suggesting that the condensate dynamics are important for reaction enhancement.

To test whether the local spatiotemporal suppression underlying these dynamics is required for $\mathcal R>0$, we replace $\chi(p)$ with its mean, $\chi_{\rm m}=\chi_0+\alpha\overline{\langle p\rangle}$.
This replacement preserves the mean suppression of condensation while removing its local dependence on $p$ and $\bm J_p^\alpha=0$. 
As $\alpha$ increases, the condensate remains static.
In this case $\mathcal R$ slightly increases, but remains negative [Fig.~\ref{fig2}(a)]. 
In this regime, mean suppression is insufficient to achieve $\mathcal R>0$; local spatiotemporal suppression by the product is required.

We then vary the production strength $\nu$ to examine how the spatial patterns and $\mathcal R$ change.
The spatial patterns are shown in Fig.~\ref{fig2}(c), and the corresponding map of $\mathcal R$ is shown in Fig.~\ref{fig6} in the Appendix.
The dependence on $\alpha$ found at $\nu=2$ is also observed over a broad range of $\nu$.
Static condensates occur at small $\alpha$.
Increasing $\alpha$ leads to rotation, rotation with deformation, and propagation, while $\mathcal R$ reaches a maximum at intermediate $\alpha$.

At sufficiently large $\alpha$ and $\nu$, suppression of condensation becomes too strong for the system to sustain a condensate.
The system then becomes homogeneous, reactant enrichment is lost, and $\mathcal R=0$.
As $\nu$ increases, the product concentration rises, so the condensate disappears at smaller $\alpha$.
Linear stability analysis yields the corresponding stability threshold $\alpha_c(\nu)$ in good agreement with the numerical results (Fig.~\ref{fig6}).

Moreover, rotating and propagating states with $\mathcal R>0$ are also obtained at $D=0$ (Fig.~S2 \cite{supp}), where both product diffusion and the coupling-induced flux $\bm J_p^\alpha$ vanish [Eq.~(\ref{eq:p_dynamics_without_D})]. 
Thus, coupling between local reaction kinetics and condensate dynamics can generate these dynamics and enhance the reaction without product transport.

\begin{figure}[t]
    \includegraphics[keepaspectratio,scale=1.0]{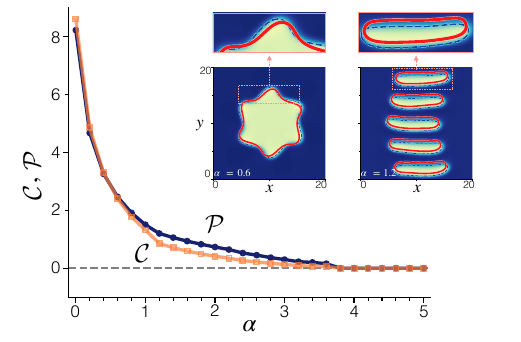}
    \caption{
    $\mathcal P$ (blue circles) and $\mathcal C$ (orange squares) versus $\alpha$.
    Insets show $\phi$ in the rotating and propagating states.
    Enlarged views of the indicated regions show the spatial offset between the contours.
    The blue dashed and red solid contours indicate $\phi=0.5$ and $p=(p_{\max}+p_{\min})/2$. 
    $D=10^{-3}$ and $\nu=2$.}
\label{fig3}
\end{figure}

To understand why $\mathcal R$ increases in the dynamic states, we examine the spatial relation between $\phi$ and $p$.
In these states, the $\phi$ and $p$ patterns are spatially shifted (Figs.~\ref{fig3} and S1~\cite{supp}).
To quantify the effect of this shift, we decompose $\mathcal R$ as
\begin{align}
\mathcal R=\mathcal P-\mathcal C,
\quad
\mathcal P=\frac{\overline{\langle p\rangle}-p_0}{p_0},
\quad
\mathcal C=\frac{
\overline{\langle\phi p\rangle}
-\phi_0\overline{\langle p\rangle}}{(1-\phi_0)p_0},
\label{eq:rate_decomposition}
\end{align}
where $\mathcal P$ measures the deviation of the mean product concentration from its homogeneous value, whereas $\mathcal C$ measures product--condensate colocalization.

Figure~\ref{fig3} shows that both $\mathcal P$ and $\mathcal C$ decrease as $\alpha$ increases and approach zero when the system becomes homogeneous.
Increasing $\alpha$ suppresses condensation and reduces $\overline{\langle\phi^2\rangle}$, causing $\mathcal P$ to decrease.
The decrease in $\mathcal C$ reflects the increasing spatial separation between the $p$-rich and $\phi$-rich regions.
At intermediate $\alpha$, $\mathcal C$ becomes smaller than $\mathcal P$, and hence $\mathcal R>0$.
This change corresponds directly to the pattern dynamics (Fig.~\ref{fig3}).
At $\alpha=0.4$, the star-shaped rotating state shows a small shift between $p$ and $\phi$, but $\mathcal R$ remains negative.
At $\alpha=0.6$, stronger deformation produces a larger shift and $\mathcal R$ becomes positive.
At $\alpha=1.2$, the propagating domains leave accumulated product behind, further reducing their overlap as $\mathcal R$ reaches its maximum.
The spatial shift therefore reduces product colocalization sufficiently to enhance the reaction despite the decrease in $\mathcal P$.

A stationary condensate cannot maintain this shift.
Because the net product production $r=r_{\rm prod}-r_{\rm conv}$ increases with $\phi$, $\partial_\phi r>0$, product accumulates wherever $\phi$ remains high and increases their colocalization.
Maintaining a small $\mathcal C$ therefore requires continual reorganization of the condensate toward product-poor regions, leaving accumulated product behind.

\begin{figure}[t]
    \includegraphics[keepaspectratio,scale=1.0]{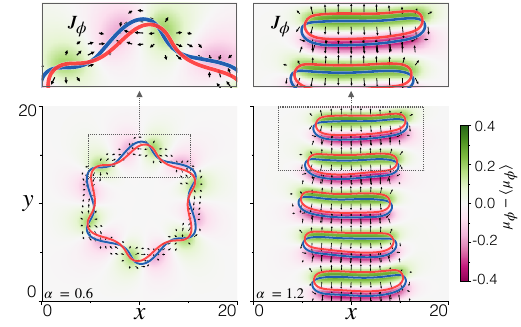}
    \caption{Chemical potential $\mu_\phi-\langle\mu_\phi\rangle$, (color) and total $\phi$ flux $\bm J_\phi$ (arrows).
    The blue and red contours indicate $\phi=0.5$ and $p=(p_{\max}+p_{\min})/2$, respectively.
    The upper panels show close-up views of the boxed regions.
    Arrows with $|\bm J_\phi|>0.1$ are shown at every third and fifth grid point for $\alpha=0.6$ and $1.2$, respectively.
    $D=10^{-3}$ and $\nu=2$.}
\label{fig4}
\end{figure}

\begin{figure}[H]
\includegraphics[keepaspectratio,scale=0.95]{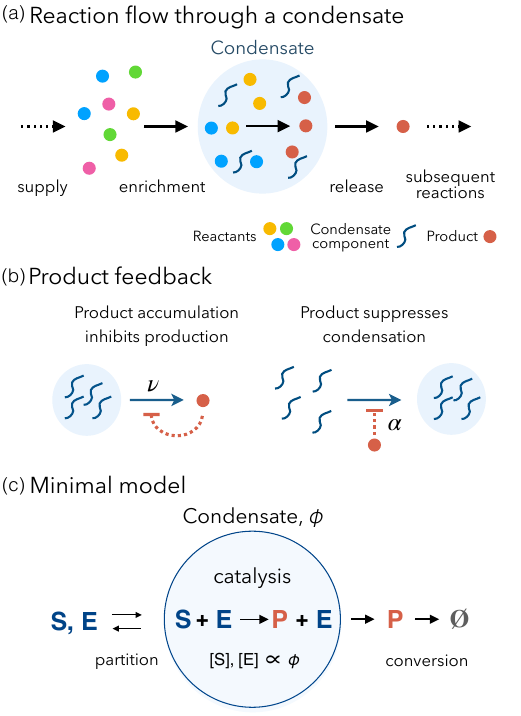}
\caption{(a) Reactants are supplied from the environment and enriched in a condensate, where they are converted into product.
The product is released and undergoes subsequent reactions.
(b) Product accumulation inhibits further production and suppresses condensation.
The parameters $\nu$ and $\alpha$ denote the production strength and the strength of product-dependent condensation suppression, respectively.
(c) Minimal realization of the reaction scheme.
A substrate $S$ and an enzyme $E$ partition into the condensate, with $[S],[E]\propto\phi$, and react as $S+E\rightarrow P+E$.
The product is subsequently converted in the dilute phase.}
\label{fig5}
\end{figure}

This continual reorganization results from the competition between product-dependent suppression and phase separation in the total $\phi$ flux \footnote{The same dynamic states are obtained at $D=0$, where $\bm J_p=0$, so we omit contribution of $\bm J_p$.}, $\bm J_\phi=-\nabla\mu_\phi^{(0)}-\alpha\nabla(p\phi)$ [Fig.~\ref{fig4}].
A local increase in $p$ raises the product-dependent contribution $\alpha p\phi$ to the chemical potential and drives $\phi$ away from that region through the flux $-\alpha\nabla(p\phi)$.
The product-independent contribution $\mu_\phi^{(0)}$ favors phase separation.
Because $\phi$ is conserved, the displaced $\phi$ condenses in a neighboring product-poor region, shifting the reaction domain [Fig.~\ref{fig4}].
Product then accumulates at the new position and induces another redistribution of $\phi$, sustaining rotation or propagation.
Increasing $\alpha$ strengthens the product-dependent flux, producing larger deformations and multiple propagating domains instead of a single rotating domain.

The condensate thus acts as a self-organized dynamic reactor.
Its continual reorganization sustains a spatial shift between the reaction domain and the accumulated product while maintaining reactant enrichment.
This dynamic separation reduces product inhibition and enhances the reaction.
Condensate dynamics and reaction enhancement persist at $D=0$, identifying local reaction--phase-separation feedback, rather than product diffusion, as their origin.

Coupling chemical reactions to phase separation can arrest coarsening or produce growth-and-division dynamics \cite{glotzer1995reaction, christensen1996phase, zwicker2017growth, wurtz2018chemical}.
Phase separation can also promote oscillations in biochemical negative-feedback loops \cite{lucchetti2026phase}, while reaction-generated substrate gradients can drive condensate motion \cite{demarchi2023enzyme, goychuk2024selfconsistent} and, through nonreciprocal chemotactic transport can generate propagating and rotating condensate patterns \cite{rasshofer2025capillary}.
In our model, by contrast, both diffusive fluxes derive from the same free energy, and the product changes the local stability of the condensed phase, causing the reaction domain to reorganize relative to the product it generates.
By considering both production and subsequent conversion, we connect this reorganization to the entire reaction flow and show that it can enhance the overall reaction rate.
Characterizing the transitions among static, rotating, deformed, and propagating condensates will require analysis of both interfacial instabilities \cite{goychuk2024selfconsistent, rasshofer2025capillary} and bifurcations in pattern formation \cite{cross2009pattern}, including how mode symmetry and nonlinear interactions generate propagation and other dynamics \cite{nishide2022pattern, nishide2025weakly}.

The effects of condensates on reactions have largely been studied under static conditions, although condensates in cells are continually assembled, dissolved, and remodeled in space and time \cite{hyman2014liquidliquid, banani2017biomolecular, lyon2021aframework, lim2024phase}.
Our results suggest that such spatiotemporal reorganization can itself regulate reactions rather than merely accompany them.
Because condensate composition regulates enzymatic activity and product partitioning \cite{harris2024regulation}, biological condensates may exploit this mechanism to control reactions.
The same principle could guide the design of synthetic reactors \cite{aierken2026roadmap}.
Rotation and propagation could provide distinct forms of spatial control in both cells and prebiotic systems, maintaining localized reaction activity or redistributing it among multiple condensate domains, respectively. 
At the origin of life, such feedback could have allowed reaction maintenance and condensate division to emerge together, providing a possible route toward protocell reproduction \cite{zwicker2017growth, kamimura2010reproduction}.
Open reaction flows can transform condensates governed by equilibrium molecular interactions into self-organized dynamic reactors.

\section*{Acknowledgement}
This study was supported by the Novo Nordisk Foundation Grant No.~NNF21OC0065542.

\bibliography{bib}

\section*{Appendix}
\vspace{-1em}
\subsection{Reaction flow and minimal model}

\subsection{Homogeneous state}
In the homogeneous state, $\phi=\phi_0$ and $p=p_0$, where $p_0$ is determined by $r_{\rm prod}=r_{\rm conv}$:
\begin{equation}
p_0=\frac{K_P}{2}\left[-1+\sqrt{1+\frac{4\nu\phi_0^2}{K_P(1-\phi_0)}}\right].
\label{eq:p0}
\end{equation}

\subsection{Reaction enhancement and homogeneous-state stability in the $(\alpha,\nu)$ plane}

\begin{figure}[H]
\includegraphics[keepaspectratio,scale=0.95]{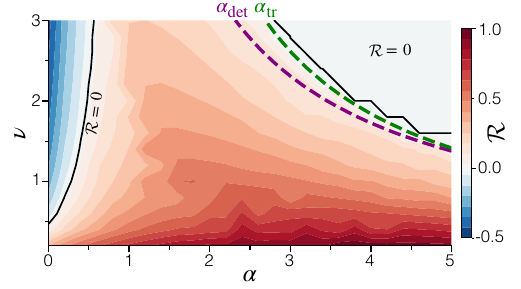}
\caption{$\mathcal R$ in the $(\alpha,\nu)$ plane at $D=10^{-3}$.
The black contour denotes $\mathcal R=0$.
The green and purple dashed curves show
$\alpha=\alpha_{\rm det}(\nu)$ and
$\alpha=\alpha_{\rm tr}(\nu)$, respectively.}
\label{fig6}
\end{figure}

Figure~\ref{fig6} shows $\mathcal R$ in the $(\alpha,\nu)$ plane at $D=10^{-3}$.
For each $\nu$, $\mathcal R$ increases with $\alpha$, reaches a maximum at intermediate $\alpha$, and then decreases at larger $\alpha$.
At small $\nu$, $\mathcal R$ is already positive at $\alpha=0$, but finite $\alpha$ increases it further.
Increasing $\nu$ generally decreases $\mathcal R$ because it increases $R_{\rm hom}$ and strengthens product inhibition through higher production in $\phi$-rich regions.

The homogeneous state is unstable for
\begin{equation}
\alpha<\alpha_c(\nu),
\quad
\alpha_c(\nu)=
\max\{\alpha_{\rm tr}(\nu),\alpha_{\rm det}(\nu)\},
\label{eq:homogeneous_instability}
\end{equation}
where
\begin{align}
\alpha_{\rm tr}
&=-\frac{2\sqrt{-\kappa\partial_p r}
+f_\phi''(\phi_0)+\chi_0+D}{p_0},\\
\alpha_{\rm det}
&=
\frac{[f_\phi''(\phi_0)+\chi_0]\partial_p r}{\phi_0\partial_\phi r-p_0\partial_p r},
\qquad (D\ll1).
\end{align}
Here, $p_0$ is the homogeneous product concentration, and the derivatives of
$r=r_{\rm prod}-r_{\rm conv}$ are evaluated at the homogeneous solution.
At $\alpha=0$, $\phi$ decouples from $p$, and the $\alpha_{\rm det}$ branch reduces to the spinodal condition $f_\phi''(\phi_0)+\chi_0<0$.
The threshold $\alpha_{\rm tr}$ is obtained by maximizing $\operatorname{tr}L(q)$ over $q$.
The resulting $\alpha_c(\nu)$ agrees with the numerical boundary of the homogeneous state.
The full derivation, including the finite-wave-number instability at finite $D$, is given in the Supplemental Material~\cite{supp}.

\end{document}